\documentclass[a4paper,12pt]{article}
\usepackage{cite}
\newcommand{\be}{\begin{equation}}
\newcommand{\ee}{\end{equation}}
\newcommand{\bea}{\begin{eqnarray}}

\newcommand{\eea}{\end{eqnarray}}

\begin{document}

\title{
\begin{flushright}
{\small WITS-CTP-030}\\
\end{flushright}\vspace{1.5cm}
An Implication of ``Gravity as the Weakest Force''}

\author{
A. J. M. Medved \\ \\
Department of Physics and Centre for
Theoretical Physics, \\
University of the Witwatersrand, \\
Wits, 2050, \\
South Africa \\
E-mail: allan.medved@wits.ac.za \\ \\} 

\maketitle
\begin{abstract}

\par
The negative specific heat of a radiating black hole
is indicative of a cataclysmic endpoint to
the evaporation process. 
In this letter, we suggest a simple mechanism for
circumventing such a dramatic outcome. The
basis for our  argument is  a conjecture
that was  recently proposed by Arkani-Hamed and collaborators.
To put it another way, we use their notion
of ``Gravity as the Weakest Force'' as a means
of inhibiting the process of black hole
evaporation.
\end{abstract}
\newpage
\section*{}
\par
It has long been realized that black holes~\footnote{We will be working
 --- for simplicity --- with 
Schwarzschild 
black holes in a four-dimensional spacetime. Nonetheless,
the discussion  easily generalizes (given a healthy
dose of algebra) into  other dimensionalities and
many classes  of  black holes.} 
 are
thermodynamic entities  with a readily 
identifiable temperature   
\be
T \;=\; {\hbar\over 8\pi GM}
\label{1}
\ee
and  entropy
\be
S \;=\; {4\pi G M^2 \over \hbar}\; ; 
\label{2}
\ee
also known as the Hawking temperature 
and the Bekenstein--Hawking entropy \cite{Beken-1,Hawk-1}.
Here, $M$ is the mass of the black hole (or, alternatively,
its Schwarzschild radius divided  by $2G$) and all non-explicit
fundamental constants have  conveniently been set to
unity. \\
\par
As an immediate consequence of this captivating framework,
black holes possess a negative specific heat ---
\be
{dM\over dT} \;=\; -{8\pi G M^2 \over \hbar} \;<\; 0 \;
\label{3}
\ee
--- meaning that the black hole must become progressively hotter
as its mass diminishes.  If one extrapolates this trend to
zero mass, the temperature diverges and the ultimate outcome would
 appear to be  an apocalyptic-like explosion. \\
\par
This last conclusion is, however, rather naive. Semi-classical 
intuitions can only be extended so far, and one expects 
quantum-gravitational effects to intervene at somewhere
around the Planck scale ({\it i.e.}, near lengths of order
$L_p=\sqrt{\hbar G}$ or masses of order $M_p=\sqrt{\hbar/G}\;$).
On the other hand, it is hard to envision how quantum gravity
can step in at (close to) the proverbial last moment 
 and suddenly halt what
is essentially a runaway process. Which is to say, one might 
anticipate some type of semi-classical ``braking" mechanism to be
 in order. 
This would presumably
 act to
(at least) slow down
the  rate of black hole radiation; at a time well before 
a Planck-order mass  has been attained. Finally,
upon reaching the Planck scale, the slowed-down rate 
of evaporation would  allow for a smooth transition from
the semi-classical to  
the quantum-gravitational realms. In this way, quantum gravity
can gracefully take over and do whatever it does to
avoid a final singularity.~\footnote{Let us suppose that there
is no braking mechanism; then  there would have
to be an abrupt phase transition
between the two realms.
This is, of course, not unfeasible but would be aesthetically
undesirable.}  \\ 
\par
So let us restrict ourselves to semi-classical physics and
then make the natural query: what braking mechanism?!
For this question, there is no 
convincing nor
definitive answer.~\footnote{The
idea  that black hole radiation
could stop prematurely  does have a history
({\it e.g.}, \cite{Gidd,Russ-1}); with the primary
motivation being its potential  to resolve 
the black hole {\it information loss} paradox \cite{Hawk-2,Russ-2}. 
 We would argue, however,
that a precise account of  just how this could happen
is still lacking. Moreover, we will contend  (near the
end of the letter) that, in deciding between  a fully stopped
and a slowed-down evaporation, the latter scenario
is actually the preferable one.}
But we can  help fill in this gap by way of an
observation that follows from a  recent conjecture  
put forward by Arkani-Hamed,  Motl, Nicolis and Vafa 
(henceforth, to be
known as AMNV)
 \cite{AMNV}.
Before detailing our observation, let us  briefly discuss the salient
 points
of the cited work.~\footnote{For related discussion, the reader
may also refer to \cite{X-1,X-2,X-3,X-4,X-5} and, especially,
\cite{Banks-1}.} \\
\par
AMNV actually make two conjectures of relevance, with
the second (the one of current interest) being a
direct consequence of the first:
\\
\par\noindent
{\it Conjecture 1}: Given a  U(1) gauge coupling with a coupling
strength of $g$, there must exist a ``sufficiently light"~\footnote{As
carefully stipulated in \cite{AMNV}.}
charged particle such that the mass ($m$) of this
particle satisfies
\be
m\;<\;g\cdot M_p \;.
\label{4}
\ee
\\
{\it Motivation for 1}: The existence of such a particle
ensures that extremal
black holes~\footnote{That is,
maximally charged  black
holes for which $M=QM_p$ (with $Q$ being the gauge charge
of the black hole in units of $g$).}
 are able to
decay and thus avoids the problems that are inherent to
the stability of highly entropic objects. To elaborate,
 if such black holes were indeed stable, there
would  be an extremely large entropy  [$\;\ln(1/g)\;$
with typically $g<<1$] associated
 with a near Planck-sized object. In this event, one
would expect {\it virtual} extremal black holes to dominate every
conceivable
scattering process, and we would have a catastrophe that is
tantamount to the so-called black hole remnant problem
\cite{Gidd,Suss-1}. [In the conventional remnant problem,
the highly entropic but stable objects would be the
conjectural entities that survive after black hole evaporation
has (prematurely) terminated.] 
\\
\par \noindent
{\it Conjecture 2}: An {\it effective} field theory that is 
relevant to
the coupling $g$ must break down at (or below) an energy scale
$\Lambda$ such that 
\be
\Lambda \;< \; g\cdot M_p \;.
\label{5}
\ee 
\\
{\it Motivation for 2}: Conjecture 2 follows from conjecture
1 by way of the following heuristic argument.~\footnote{The
current author finds the second conjecture to be somewhat less
compelling than the first. Nevertheless,  the AMNV
paper has stirred sufficient interest to merit a treatment
such as ours, irrespective of any given author's or reader's
personal opinion.}  For a magnetic monopole
(or an analogue thereof),  equation (\ref{4}) may be recast
as
\be
m\;<\; {1\over g}\cdot M_p \;.
\label{6}
\ee
Meanwhile, if the field theory has an ultraviolet cutoff
of $\Lambda$, then the (otherwise divergent) monopole mass is
expected to be of the order 
\be
m\;\sim\;{\Lambda\over g^2}  \;.
\label{7}
\ee
Combining equations (\ref{6}) and (\ref{7}), one finds
that the bound of equation (\ref{5}) trivially follows.
Let us also mention here that the AMNV  notion of ``gravity as the 
weakest
force" follows from  the gravity force
(with ``coupling" $m$) being overwhelmed by the gauge force 
 when expressed in Planck units ($\;m<g\;$).
\\
\par
We will now, as advertised, proceed to exploit the  AMNV 
conjectures  (in particular, the second one)
in the context of black hole
evaporation.~\footnote{By agreeing to work with
the Schwarzschild model, we are really insisting
that the black hole is (essentially) neutral with respect to
all  gauge charges that are implied by 
this discussion. Fortunately, from a realist's perspective,
such neutrality would be the normal state of
affairs.}
 Let us remind the reader that the discussion
is to be kept at the level of semi-classical physics,
so that conventional quantum-mechanical reasoning and
effective field-theoretic descriptions should both apply.
\\
\par
First of all, 
a necessary prerequisite for a particle to be radiated 
by a black hole is the capability of ``fitting inside". 
To elaborate, let us suppose that some particle spatially extends far 
outside the black 
hole (and we  will appropriately use its Compton wavelength as a
 measure of this
extent), while a second particle is localized entirely
inside the horizon. Then, by simple probability arguments, the
former (delocalized) particle has much  less chance of 
interacting with the black hole  gravitational field.
(Obviously, such an interaction much precede the process of radiation.) 
It thus follows that
the Compton length
of a radiated particle should be (roughly) bounded by the 
Schwarzschild 
radius of
the black hole,~\footnote{Alternatively, 
this is
just a  restatement of the 
following fact \cite{Hawk-1}:  
the major contribution to Hawking radiation comes from
 particles with wavelengths that are peaked around the inverse of the
 Hawking
 temperature --- that is, peaked about the Schwarzschild radius
[{\it cf}, equation(\ref{1})].}
 or~\footnote{One might be concerned over the
values inserted into equation (\ref{8}), given that length (energy)
scales
are  extremely
red (blue) shifted in the proximity of a black hole
horizon.  For our discussion, however, the relevant observer is  
asymptotically situated,
 as this is the type of observer who would detect  
thermal emissions from a black hole. 
Significantly, such an observer would attribute the lengths
in question with the values as given in equation (\ref{8}).}
\be
{\hbar\over m}\;<\; 2 G M\;.
\label{8}
\ee
Next, applying $m<\Lambda$ (as must be the case given
an effective field-theoretical description) and $M_p^2=\hbar/G$,
 we have
\be
{1\over \Lambda}\;<\; {M\over M_p^2} \;.
\label{9}
\ee
\\
\par
We will now bring the gravity-as-the-weakest-force conjecture
in the guise of equation (\ref{5}) into play. The inversion
$\;1/gM_p\;<\; 1/\Lambda\;$ allows
us to rewrite equation (\ref{9}) in the following way:
\be
{1\over g} \;<\;{M\over M_p} \;
\label{10}
\ee
or, more succinctly, in Planck units,
\be
{1\over g} \;<\; M \;.
\label{11}
\ee 
\\
\par
We have finally arrived at the crux of the matter.
As the black hole radiates away particles, $M$ obviously becomes
progressively smaller. Hence, as the evaporation proceeds, the bound on 
$1/g$ becomes tighter and tighter. (Recall that, typically,
$\;g<<1\;$,  so that  $\;1/g>>1\;$.) Meaning, the number of species
of particles that  the black hole can emit is
monotonically decreasing throughout the evaporation.
To put it another way, light-massed black holes
will only be able to emit the most strongly interacting
(and, presumably, rarest) species of particles.
Hence, we have a natural mechanism that --- not only inhibits the black
hole radiative process but --- works at direct cross purposes
to the destabilizing effect of the negative specific heat.
 Let us suppose that this species-drop-off effect is sufficiently
competitive with the (otherwise) accelerating evaporation;
sufficient enough so that
the black hole can remain
relatively stable upon its approach to the Planck scale.
Then we have identified  precisely the type
of braking mechanism that was advocated earlier in the letter.
\\
\par
One might wonder if our rigorous restriction
 to a semi-classical framework is too stringent a condition.
After all, the immense gravitational  attraction of a black
hole coupled with the intrinsically 
quantum process of it radiating is strongly 
suggestive of a quantum-gravitational realm.
But here is the vital point: The process of
black hole radiation is, as we best understand it,
 based solely
on semi-classical reasoning.~\footnote{Having said that,
we should acknowledge the progress made in 
understanding black hole radiation in the context of
string theory \cite{Peet} and loop quantum gravity \cite{Ash}.
These  are, however, highly model-specific methodologies.}
 Most notably, Hawking's
original derivation \cite{Hawk-1} in the
context of (``standard") quantum field theory in curved spacetime
\cite{BD}; although other semi-classical treatments
have deduced black hole thermality with the same value
for the temperature ({\it e.g.}, \cite{Y-1,Y-2,Y-3,Y-4,Y-5}).
Hence, once quantum gravity is explicitly needed in 
 the discussion, we have no compelling
reason to believe that the  Hawking radiative process
survives in any recognizable form.~\footnote{Interestingly,
the same basic rationale has been used to argue {\it generically}
against
black hole  radiation  \cite{Helf}.
The quantum-gravitational sticking point 
is, in this case,  the trans-Planckian energies
that unavoidably arise when one traces the outgoing particles back
to the horizon.}
\\
\par
To summarize, we have argued that the hypothesis of
 ``gravity as the weakest force"
(as conjectured  by Arkani-Hamed {\it et. al.} \cite{AMNV})
implies a mechanism that competes directly with
the negative specific heat of a shrinking black hole.
This mechanism ---  which is based on limiting the number
of  particle species available for black hole emission ---
should be able to slow down an otherwise runaway evaporation
process. Assuming this braking is strong enough, we have
anticipated a smooth transition from the semi-classical
to the quantum-gravitational
regimes as the black hole mass descends towards 
the Planck scale.
\\
\par
Even more ambitiously, our
conjectured mechanism could halt the process of
black hole evaporation 
altogether; thus   resolving the conundrum of
what happens to the information that was trapped inside
of a black hole after it evaporates --- the so-called
information loss paradox \cite{Hawk-2,Gidd,Russ-2}.
On the other hand, because of the  previously
mentioned remnant problem 
({\it i.e.}, the existence
 of small stable objects with extremely large entropy 
would wreak havoc on low-energy physics), it
is not clear that this is a favorable resolution. In fact,
given the context of the current discussion, it is clearly
{\it not} favorable. That is, if the evaporation stopped
completely, then the AMNV conjecture would simply
be trading off one remnant-type problem 
(extremal black holes)
 for another (Schwarzschild remnants).
\\
\par
 As a final note, let us point out
that such a braking mechanism provides the potential for
a experimentally verifiable signature of
the gravity-as-the-weakest-force paradigm.

\section*{Acknowledgments}
Research is supported by 
NRF research grant Gun 2053791 and a URC Postdoctoral Fellowship.
The author thanks Robert de Mello Koch for
his inputs and inspiring discussions.


\end{document}